\documentclass[preprint,aps,eqsecnum,superscriptaddress]{revtex4}
\usepackage{amsfonts}
\usepackage{epsfig}
\begin{document}

\title{Noncommutative Field Theory: Nonrelativistic Fermionic Field  
Coupled to the Chern-Simons Field in 2+1 Dimensions}
\author{M. A. Anacleto}
\author{M. Gomes}
\author{A. J. da Silva}
\email{anacleto, mgomes, ajsilva@fma.if.usp.br}
\author{D. Spehler}
\altaffiliation[On leave from] {Institut de Recherches Subatomiques, in2p3/CNRS and Universit\'e Louis Pasteur, Strasbourg, France}
\email{spehler@if.usp.br}
\affiliation{Instituto de F\'\i sica, Universidade de S\~ao Paulo\\
Caixa Postal 66318, 05315-970, S\~ao Paulo, SP, Brazil}
\date{\today}
\begin{abstract}
\pretolerance1000
We study a noncommutative nonrelativistic fermionic field theory in 2+1
dimensions  coupled to the Chern-Simons field. We perform a perturbative
analysis of model  and  show that up to one loop the ultraviolet
divergences are canceled and the infrared divergences are eliminated
by the noncommutative Pauli term.
\end{abstract}
\maketitle
\pretolerance1000

\section{Introduction}

The noncommutative Aharonov-Bohm (AB) effect for scalar particles has
been studied in the quantum mechanical \cite{chaichian,Gamboa} and 
in the field theory contexts \cite{anacleto}.  
As in the commutative case, in the latter case the effect
was simulated by a nonrelativistic field theory of spin zero particles
interacting through a Chern-Simons (CS) field. Differently from its
commutative counterpart, however, the model turns out to be renormalizable even
without a quartic self-interaction of the scalar field (the quartic
self-interaction is however necessary if a smooth commutative limit is
required).  It is known, that in the commutative situation, the Pauli
term plays for the spin $1/2$ AB scattering \cite{Gomes} the same role as the
quartic interaction plays for the case of scalar particles \cite{esc}. One may
then conjecture that a noncommutative Pauli interaction is also not 
necessary at
least as a pre-requisite for renormalizability.
In this brief note we will prove that this conjecture indeed holds,
the Pauli term being necessary to obtain a smooth result in the commutative
limit but not to fix the ultraviolet renormalizability of the model.
Our analysis is based in the (2+1) dimensional model
described by the action
\begin{eqnarray}
\label{accao} 
S[A,\psi ] &=&\int d^{3}x\left[ \frac{\kappa }
{2}\varepsilon ^{\mu \nu\lambda }
\left( A_{\mu }\partial _{\nu }A_{\lambda }
+\frac{2ig}{3}A_{\mu }\ast A_{\nu }\ast A_{\lambda }\right)
-\frac{1}{2\xi }\partial_{i}A^{i}\partial _{j}A^{j}
+i\psi ^{\dagger }\ast D_{t}\psi\right. 
\nonumber\\
&&-\left.\frac{1}{2m}({\bf D}\psi)^{\dagger }\ast ({\bf D}\psi )
+\lambda\psi^{\dagger}\ast B\ast\psi 
+\partial ^{i}\bar{c}\partial _{i}c
+ig\partial ^{i}\bar{c}\ast[A_{i},c]_\ast\right],
\end{eqnarray}
where $B=-F_{12}=\partial_{1}A^{2} -
\partial_{2}A^{1}+ig[A^{1},A^{2}]$ 
and $\psi$ is a one-component fermion field transforming in accord
with the fundamental representation of the noncommutative $U(1)$
group (other noncommutative aspects of nonrelativistic fermions
interacting with the CS field were considered in \cite{Horvathy,Ghosh})

\begin{equation}
\psi\rightarrow (e^{i\Lambda})_{\ast}\ast\psi, 
\end{equation}
\begin{equation}
\psi^{\dagger}\rightarrow \psi^{\dagger}\ast(e^{-i\Lambda})_{\ast}, 
\end{equation}
The covariant derivatives in Eq. (\ref{accao}) are given by 
\begin{eqnarray}
D_{t}\psi &=&\partial_{t}\psi+igA_{0}\ast\psi,  \nonumber\\
D_{i}\psi &=&\partial _{i}\psi+igA_{i}\ast\psi.
\end{eqnarray}

\begin{figure}
\centering
\scalebox{0.7}{\includegraphics{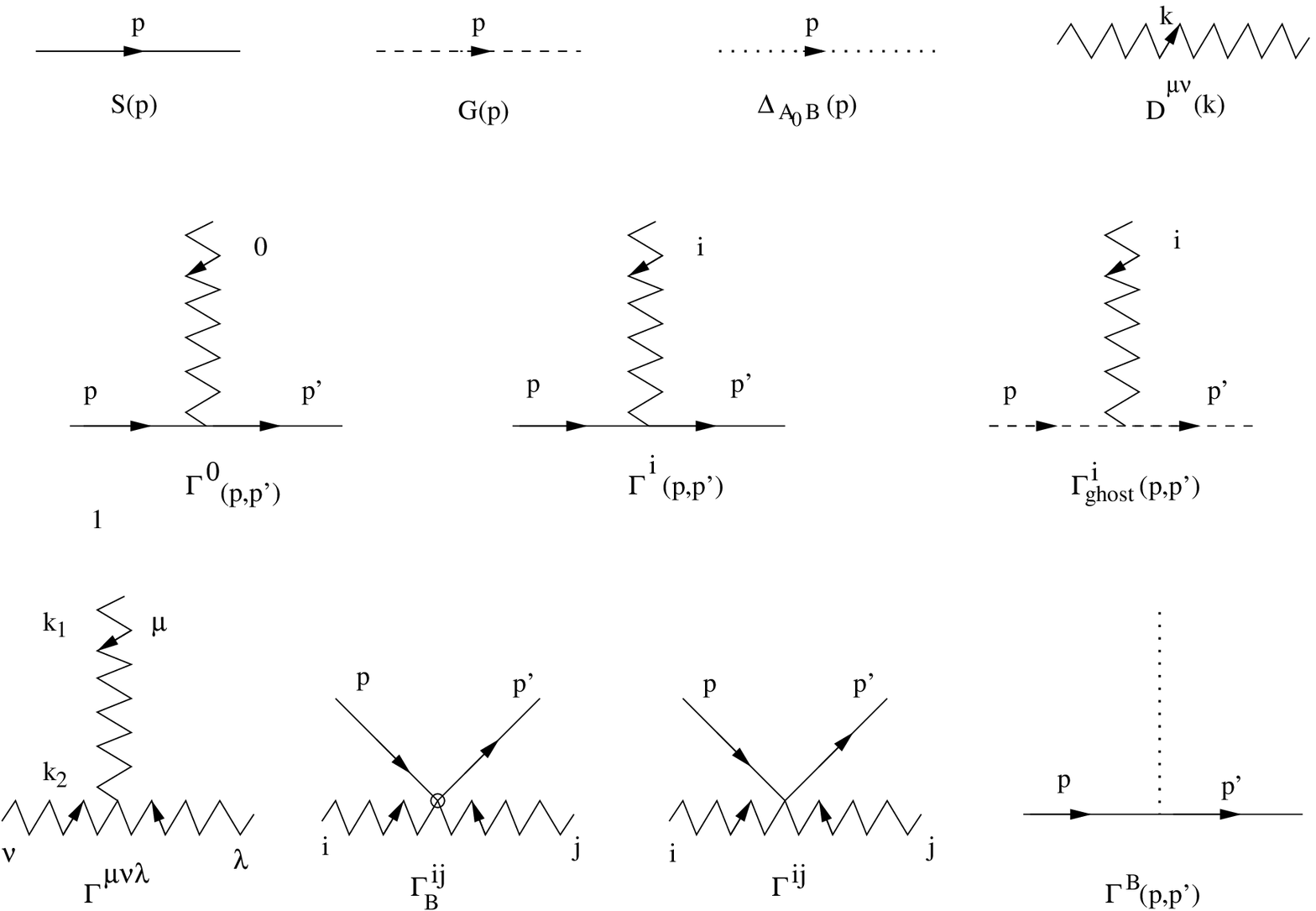}}
\caption{Feynman rules for the action (\ref{accao}).}
\label{regrab}
\end{figure}

\noindent
For convenience, we will work in a strict Coulomb gauge obtained
by letting $\xi \rightarrow 0$.
We will use a graphical notation where the CS field, the
matter field and the ghost field propagators are represented by
wavy, continuous and dashed line respectively.
The graphical representation for the Feynman rules is given
in Fig. \ref{regrab} and the corresponding analytical expression are:

\noindent
(i) The matter field propagator:
\begin{equation}
S(p)=\frac{i}{p_{0}-\frac{{\bf p}^2}{2m} + i\epsilon},
\end{equation}

\noindent
(iii) The ghost field propagator: 
\begin{equation}
G(p)=-\frac{i}{{\bf p}^2},
\end{equation}

\noindent
(iii) The mixed propagator:
\begin{equation}
\Delta_{A_{0}B}(p)=-\frac{i}{\kappa},
\end{equation}

\noindent
(iv) The gauge field propagator in the limit $\xi \rightarrow 0$ is 
\begin{equation}
D_{\mu \nu }(k)=\frac{\varepsilon_{\mu\nu\lambda}\bar{k}^{\lambda}}
{\kappa {\bf k}^{2}},
\end{equation}
where $\bar{k}^{\lambda}$=$(0,{\bf k})$.

\noindent
(v) The analytical expressions associated to the vertices are:
\begin{eqnarray}
\Gamma ^{0}(p,p^{\prime}) &=&-ige^{ip\theta p^{\prime}}, \\
\Gamma ^{i}(p,p^{\prime})&=&\frac{ig}{2m}(p+p^{\prime})^{i}
e^{ip\theta p^{\prime}}, \\
\Gamma ^{i}_{ghost}(p,p^{\prime})&=&-2gp^{\prime{i}}
\sin(p\theta p^{\prime}), \\
\Gamma ^{\mu\nu\lambda}(k_{1},k_{2}) &=&2ig\kappa \varepsilon ^{\mu \nu
\lambda }\sin (k_{1}\theta k_{2}), \\
\Gamma^{ij}_{B}(k_{1},k_{2},p,p^{\prime})&=&2ig\lambda\sin(k_{1}\theta k_{2})
e^{ip\theta p^{\prime}}\varepsilon^{ij}, \\
\Gamma ^{ij}(k_{1},k_{2},p,p^{\prime}) &=&-\frac{ig^{2}}{m}
\cos(k_{1}\theta k_{2})e^{ip\theta p^{\prime}}\delta ^{ij},\\
\Gamma^{B}(p,p^{\prime})&=&i\lambda e^{ip\theta p^{\prime}}. 
\end{eqnarray}
In these expressions we have defined
$k_{1}\theta k_{2}\equiv \frac{1}{2}\theta ^{\mu\nu}k_{1\mu}k_{2\nu}$,
where $\theta^{\mu\nu}$ is the antisymmetric matrix which characterizes
the noncommutativity of the underlying space.
For simplicity and to evade possible
unitarity/causality problems \cite{Gomis} we keep time local by imposing
$\theta^{0i}=0$. We set also $\theta^{ij}= \theta \varepsilon^{ij}$ with
$\varepsilon^{ij}$ being the two dimensional Levi-Civit\`a symbol,
normalized as $\varepsilon^{12}=1$.

\section{Particle-Particle Scattering}

\subsection{Tree Level Scattering}

\begin{figure}
\centering
\scalebox{0.7}{\includegraphics{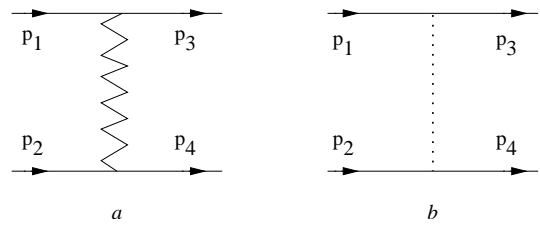}}
\caption{Tree level scattering.}
\label{treelevel}
\end{figure}

In the tree approximation and in the center-of-mass frame the two body
scattering amplitude, depicted in Fig. \ref{treelevel}$a$, is given by
\begin{equation}
\label{espa}
{\cal{A}}_{a}^{0}(\varphi)=-\frac{2ig^2({\bf p}_{1}\wedge {\bf p}_{3})}
{m\kappa}\left[\frac{e^{i(p_{1}\theta p_{3}+p_{2}\theta p_{4})}}
{({\bf p}_{1}-{\bf p}_{3})^{2}}
-\frac{e^{-i(p_{1}\theta p_{3}+p_{2}\theta p_{4})}}
{({\bf p}_{1}+{\bf p}_{3})^{2}}\right],
\end{equation}
where ${\bf p}_{1}$, ${\bf p}_{2}$ and ${\bf p}_{3}$,
${\bf p}_{4}$ are, respectively, the incoming and outgoing momenta. Since 
$\theta_{ij}=\theta\varepsilon_{ij}$, the phase is 
$p_{1}\theta p_{3}+p_{2}\theta p_{4}=\theta({\bf p}_{1}\wedge {\bf p}_{3})
=\theta{\bf p}^{2}\sin\varphi=\bar{\theta}\sin\varphi$,
where we have defined $\bar{\theta}\equiv \theta{\bf p}^{2}$ and
$\varphi $ is the scattering angle. Therefore,
Eq. (\ref{espa}) can be rewritten as 
\begin{equation}
{\cal{A}}_{a}^{0}(\varphi)=-\frac{ig^2}{m\kappa}
\left[\frac{e^{i\bar{\theta}\sin\varphi}}
{1-\cos\varphi}
-\frac{e^{-i\bar{\theta}\sin\varphi}}
{1+\cos\varphi}\right]\sin\varphi.
\end{equation}

For the graph in Fig.  \ref{treelevel}$b$, we have
\begin{equation}
{\cal{A}}_{b}^{0}(\varphi)=-\frac{4g\lambda}{\kappa}
\cos(\bar{\theta}\sin\varphi).
\end{equation}
Thus, the full tree level amplitude is
\begin{equation}
{\cal{A}}^{0}(\varphi)=-\frac{ig^{2}}{m\kappa}
[\cot(\varphi/2)e^{i\bar{\theta}\sin\varphi}
-\tan(\varphi/2)e^{-i\bar{\theta}\sin\varphi}]
-\frac{4g\lambda}{\kappa}\cos(\bar{\theta}\sin\varphi),
\end{equation}
furnishing up to first order in the parameter $\bar{\theta}$,  
\begin{eqnarray}
{\cal{A}}^{0}(\varphi)
&=&-\frac{2ig^{2}}{m\kappa}
(\cot\varphi+i\bar\theta)
-\frac{4g\lambda}{\kappa} + {\cal O}({\bar{\theta}}^{2}).
\end{eqnarray}
Notice that the noncommutative contribution is isotropic although 
energy dependent.
\subsection{One Loop Scattering}

\begin{figure}
\centering
\scalebox{0.7}{\includegraphics{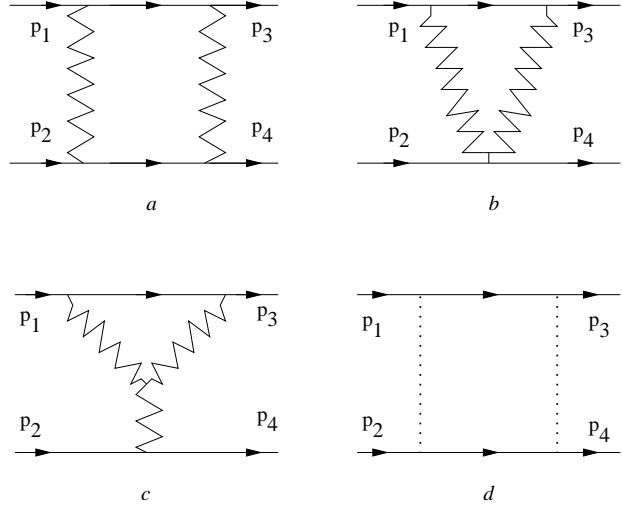}}
\caption{One-loop scattering.} 
\label{umloop}
\end{figure}

The one-loop contribution to the  scattering amplitude is depicted in 
Fig. \ref{umloop}. 
Two other diagrams corresponding to graphs \ref{umloop}$b$ and 
\ref{umloop}$c$ with the upper and bottom fermionic lines exchanged not
explicitly shower.
All other possible one-loop graphs vanish.
The expressions for  the contributions of the box, triangle and trigluon
graphs, shown in Figs. \ref{umloop}$(a-c)$, are the same as in the
scalar
case \cite{anacleto} so that we just quote the results: 
\begin{equation}
{\cal{A}}_{a}(\varphi)=-\frac{g^{4}}{2\pi m\kappa ^{2}}
\left[\ln(2\sin\varphi) +i\pi \right]
-\frac{i\bar{\theta}g^{4}\sin\varphi}{\pi m\kappa ^{2}}
\ln\left[\tan\left(\frac{\varphi}{2}\right)\right]
+ {\cal O}({\bar{\theta}}^{2}),
 \end{equation} 
for the total contribution of the box graph,
\begin{eqnarray}
{\cal{A}}_{b}^{np}(\varphi)&=&\frac{g^{4}}{2\pi m\kappa ^{2}}
[\ln(\bar{\theta}/2)+\gamma]
+\frac{g^{4}}{2\pi m\kappa ^{2}}\ln (2\sin\varphi) 
\nonumber\\
&&+\frac{i\bar{\theta}\sin\varphi g^{4}}{2\pi m\kappa ^{2}}
\ln[\tan(\varphi/2)]+ {\cal O}({\bar{\theta}}^{2})
\end{eqnarray}
and
\begin{eqnarray}
{\cal{A}}_{b}^{p}(\varphi)&=&-\frac{g^{4}}{4\pi m\kappa ^{2}}
[\cos(\bar{\theta}\sin\varphi)\ln \left(\frac{\Lambda ^{2}}
{{\bf p}^{2}}\right) \nonumber\\
&&-\ln |2\sin (\varphi /2)|e^{i\bar{\theta}\sin\varphi}
-\ln|2\cos (\varphi /2)|e^{-i\bar{\theta}\sin\varphi}],
\end{eqnarray}
for the  nonplanar and planar parts of the triangle graph,
\begin{eqnarray}
{\cal{A}}_{c}^{np}(\varphi)&=&\frac{3g^{4}}{2\pi m\kappa ^{2}}
[\ln(\bar{\theta}/2)+\gamma]
+\frac{3g^{4}}{2\pi m\kappa ^{2}}\ln (2\sin\varphi) 
\nonumber\\
&&+\frac{3i\bar{\theta}\sin\varphi g^{4}}{2\pi m\kappa ^{2}}
\ln[\tan(\varphi/2)]
+\frac{2g^{4}}{\pi m\kappa ^{2}}+ {\cal O}({\bar{\theta}}^{2})
\end{eqnarray}
and
\begin{eqnarray}
{\cal{A}}_{c}^{p}(\varphi)&=&\frac{g^{4}}{4\pi m\kappa ^{2}}
[\cos(\bar{\theta}\sin\varphi)
\left[\ln\left(\frac{\Lambda ^{2}}{{\bf p}^{2}}\right) +1\right]
\nonumber \\
&&-\ln |2\sin (\varphi /2)|e^{i\bar{\theta}\sin\varphi}
-\ln|2\cos (\varphi /2)|e^{-i\bar{\theta}\sin\varphi}],
\end{eqnarray}
for the  nonplanar and planar parts of the trigluon graph.

\begin{figure}
\centering
\scalebox{0.7}{\includegraphics{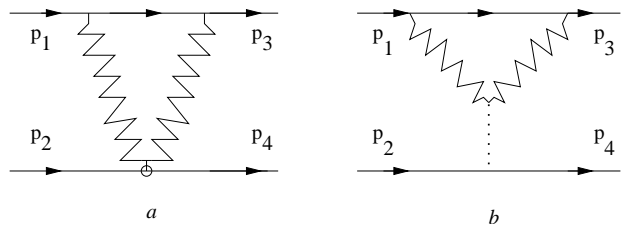}}
\caption{Pauli's term.} 
\label{grafpauli}
\end{figure}

The graphs containing the noncommutative Pauli vertex are depicted in
Fig. \ref{umloop}$d$ and Fig. \ref{grafpauli}$(a,b)$. 

The contribution of the graph in
Fig. \ref{umloop}$d$,  which is purely nonplanar, is given by 
\begin{eqnarray}
{\cal{A}}_{d}^{np}(\varphi)&=&
\frac{4mg^{2}\lambda^{2}}{\kappa^{2}}
\int\frac{d^{2}{\bf k}}{(2\pi)^{2}}
\left[\frac{e^{2iq\theta k} + e^{2iq^{\prime}\theta k}}
{{(\bf k}^{2}-{\bf p}^{2}-i\epsilon)}\right]
\end{eqnarray}
thus, by evaluating the integral in the momenta~\cite{Gel}
gives
\begin{eqnarray}
\!\!\!\!{\cal{A}}_{d}^{np}(\varphi)&=&
-\frac{4mg^{2}\lambda^{2}}{\pi\kappa^{2}}
\left[\ln\left(\frac{\bar{\theta}}{2}\right) 
+\gamma \right]
-\frac{2mg^{2}\lambda^{2}}{\pi\kappa^{2}}\ln[2\sin\varphi]
+\frac{2img^{2}\lambda^{2}}{\kappa^{2}}+{\cal O}({\bar\theta}^{2}),
\end{eqnarray}
for small $\theta$. As can be easily verified, the contributions from
the
other two graphs, shown in Figs. \ref{grafpauli}$a$ and \ref{grafpauli}$b$ 
cancel among themselves.

Summing all the contributions, we get the total one-loop amplitude
\begin{eqnarray}
{\cal{A}}_{\mbox{1-loop}}(\varphi)
&=&{\cal{A}}_{\mbox{1-loop}}^{p}(\varphi)
+{\cal{A}}_{\mbox{1-loop}}^{np}(\varphi)+{\cal{A}}_{a}(\varphi ) 
\nonumber\\
&=&-\frac{2ig^{2}}{m\kappa}\cot\varphi
-\frac{4g\lambda}{\kappa}+\frac{2\bar{\theta}g^{2}}{m\kappa}
+\frac{9g^{4}}{4\pi m\kappa ^{2}}-\frac{ig^{4}}{2m\kappa ^{2}}
+\frac{2img^{2}\lambda^{2}}{\kappa^{2}}
\nonumber\\
&&+\left(\frac{3g^{4}}{2\pi m\kappa^{2}}-\frac{2mg^{2}\lambda^{2}}
{\pi\kappa^{2}}\right)\ln [2\sin\varphi]
+\frac{i\bar{\theta}g^{4}\sin\varphi}{\pi m\kappa ^{2}}
\ln[\tan(\varphi/2)] 
\nonumber\\
&&+\frac{4mg^{2}}{\pi\kappa ^{2}}\left(\frac{g^{2}}{2m^{2}}-\lambda^{2}\right)
\left[\ln(\bar{\theta}/2)+\gamma\right]
+{\cal O}({\bar\theta}^{2}).
\end{eqnarray} 
For $\lambda=\pm\frac{g}{\sqrt{2}m}$, the limit 
$\bar{\theta}$ $\rightarrow 0$ is analytical. 
Hence, we show that the scattering amplitude up to one-loop order for
the noncommutative nonrelativistic fermionic field theory in 2+1
dimensions coupled to the Chern-Simons field does not present ultraviolet
divergences and the IR divergences in the  scattering amplitude are canceled 
by the noncommutative Pauli term. 

In this work we have studied  the two body  scattering amplitude when 
the colliding particles both have either spin up or spin down. If the
particles in colliding beams have opposite spins the contribution of
the noncommutave Pauli terms cancels. In this case, designating by $\psi$ and $\phi$
the fermionic fields associated with the particles in the beams, to
get  a smooth commutative limit it will be necessary to include
a quartic term $\phi^\dag\ast\phi\ast\psi^\dag\ast\psi$. In fact, in
the commutative situation one such term is induced if one starts
from the  relativistic theory \cite{gomes} and integrates over the
high
energy modes to get an effective nonrelativistic field theory.

\newpage
\end{document}